\documentclass[conference]{IEEEtran}
\IEEEoverridecommandlockouts
\usepackage{cite}
\usepackage{hyperref}
\usepackage{amsmath,amssymb,amsfonts}
\usepackage{listings}
\usepackage{algorithmic}
\usepackage{graphicx}
\usepackage{textcomp}
\usepackage{xcolor}
\usepackage{url}
\usepackage{placeins}
\usepackage{booktabs}

\def\BibTeX{{\rm B\kern-.05em{\sc i\kern-.025em b}\kern-.08em
    T\kern-.1667em\lower.7ex\hbox{E}\kern-.125emX}}

\begin{document}

\newcommand{\nb}[2]{
    \fbox{\bfseries\sffamily\scriptsize#1}
    {\sf\small\textcolor{red}{\textit{#2}}}
}
\newcommand{\nbc}[2]{
    \fbox{\bfseries\sffamily\scriptsize#1}
    {\sf\small\textcolor{blue}{\textit{#2}}}
}
\newcommand\ag[1]{\nb{AG}{#1}}
\newcommand\ic[1]{\nbc{IC}{#1}}
\newcommand\cb[1]{\nbc{CB}{#1}}

\newtheorem{puzzle}{Puzzle}
\newtheorem{example}{Example}

\title{Data-Driven Forward and Inverse Modeling of V-Beam Thermal Sensors}

\author{
\IEEEauthorblockN{Tudor Bartha, Adrian Groza}
\IEEEauthorblockA{\textit{Artificial Intelligence Research Institute (AIRi@UTCN}),\\ \textit{Technical University of Cluj-Napoca}}
\and
\IEEEauthorblockN{Radu Chiorean}
\IEEEauthorblockA{\textit{Faculty of Mechanical Engineering} \\
\textit{Technical University of Cluj-Napoca}}
}

\maketitle

\begin{abstract}
This paper presents a machine learning framework for data-driven inverse design of V-beam thermal sensors. 
The goal is to determine the optimal sensor geometry: beam inclination angle, beam length and beam width that achieves a target displacement under a given temperature. The design should also provide the geometry with minimum structure volume and minimum mechanical stress the sensor must support. This problem is ill-posed as for a given displacement there are multiple possible geometric configurations, causing direct regression methods to fail. We document a series of five exploratory trials that progressively revealed the nature of the problem culminating in a two-phase solution: a neural network forward model trained to map geometry and material constants to sensor responses, a gradient-descent inverse optimization over the frozen forward model, minimizing stress and volume simultaneously. The proposed pipeline utilizes a 3000-sample dataset and achieves a MAPE of 4.76\% for predicting the displacement, more than 70\% of predictions having MAPE of under 5\%.

\end{abstract}

\section{Introduction}

The V-beam thermal sensor, often referred to in MEMS literature as a V-shaped or chevron electrothermal microstructure, is a compact microelectromechanical structure that converts thermal energy into mechanical displacement or actuation force. It typically consists of inclined microbeams anchored at both ends and connected to a central shuttle. When Joule heating or an external thermal field increases the temperature of the beams, thermal expansion occurs; because the beams are geometrically constrained, this expansion is redirected into in-plane motion. This configuration provides high force density, relatively large displacement, and efficient use of chip area compared with many electrostatic MEMS structures \cite{sinclair2000high}.

V-beam MEMS structures are important because they combine electrical, thermal, and mechanical domains in a single miniaturized device. Their performance depends on beam geometry, temperature distribution, material properties and heat transfer to the surrounding environment. Furthermore, they are useful not only as actuators but also as thermal-sensitive structures in sensing and characterization platforms. Thermal MEMS devices have been widely studied for automotive systems, defense technologies, health-care systems, optical switches and nano-positioning applications \cite{phinney2012thermal}. 

The V-beam structure is especially suitable for studies involving machine learning models applied to mechanical structures because of their multiphysics problem and nonlinear dependency on the defining geometrical parameters. Recent research shows that V-shaped electrothermal MEMS devices are used in micro motors, micro robots, optical lens scanners, microgrippers and rotary or linear micromotors \cite{zhang2021macro, dzung2019iterative, li2024fabrication}. Machine learning has also been introduced to improve modelling accuracy; for example, a BP neural-network-based modelling approach reduced the error between ANSYS simulation and experimental results for a V-shaped electrothermal MEMS actuator to less than 1\% \cite{zhang2021macro}. Therefore, the V-beam thermal MEMS sensor represents a strong candidate structure for data-driven prediction of displacement, temperature, stress, reliability, and coupled electro-thermo-mechanical behavior. 

\begin{figure}[htbp]
    \centering
    \includegraphics[width=1.2\columnwidth]{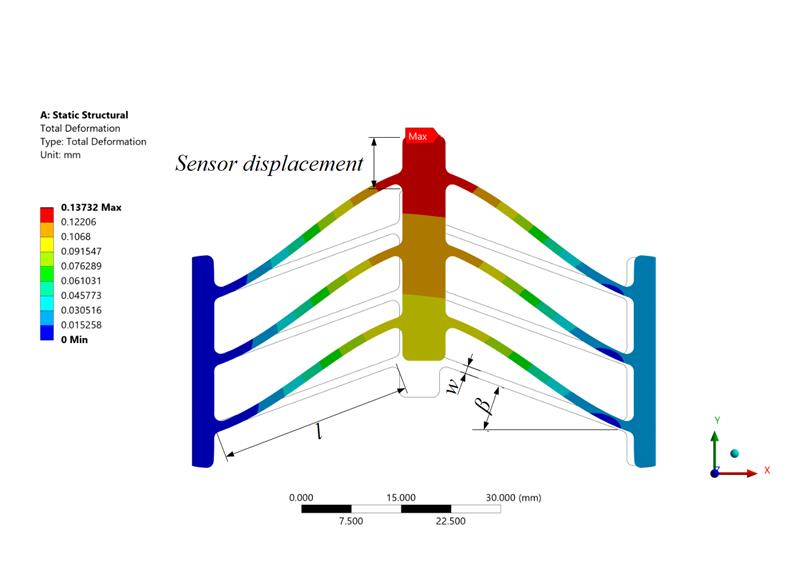}
    \caption{V-beam thermal sensor response and geometrical parameters}
    \label{fig:sensor}
\end{figure}

Figure \ref{fig:sensor} shown below illustrates the V-beam thermal sensor response and the geometrical parameters we have chosen to define the structure for the purpose of this study. The displacement is obtained using FEA simulation in ANSYS Workbench. A uniform thermal condition is applied to the structure corresponding to a sensing application with slow temperature variations. 

\section{Dataset Description}

\textit{1) Dataset Overview:} The dataset was generated using FEA simulation with Ansys. Random parameters were selected within given intervals: $\beta \in [10, 40]$ degrees, $l \in [20, 35]$ mm, $w \in [1, 2]$ mm with a step of 0.1. The dataset contains 3000 samples.

\begin{table}[htbp]
\caption{Dataset Features and Labels}
\label{tab:dataset}
\centering
\begin{tabular}{|l|l|l|l|}
\hline
\textbf{Variable} & \textbf{Role} & \textbf{Range} & \textbf{Unit} \\
\hline
$\Delta T$    & Input  & 20--70             & K \\
Displacement  & Input  & 0.025--0.2902      & $\mu$m \\
$\beta$ (beta angle)       & Output & 10--40             & degrees \\
$l$ (beam length)           & Output & 20--35             & mm \\
$w$ (beam width)           & Output & 1--2               & mm \\
Max Stress    & Output & 26.16--172.62      & MPa \\
Volume        & Output & 497--817.2         & mm$^3$ \\
\hline
\end{tabular}
\end{table}

\textit{2) Intended Use:} This dataset is intended for training and evaluating machine learning models for the inverse design of V-beam thermal sensors, specifically predicting optimal geometric parameters ($\beta$, $l$, $w$) for a target displacement under a given temperature load.

\textit{3) Out-of-Scope Use:} This dataset should not be used for MEMS geometries other than the V-beam configuration, nor for materials with properties outside the simulation parameters without retraining.

\textit{4) Data Collection and Processing:} Simulation parameters were sampled randomly within the specified intervals using Ansys Mechanical FEA simulations. No additional filtering or normalization was applied at the data collection stage; preprocessing (StandardScaler) is applied during training.

\textit{5) Bias, Risks, and Limitations:} Due to the random simulation, the dataset in not uniformly distributed. Samples with $sensor displacement \in (0, 0.15)$ constitute 2726 out of 3000 geometries. Since no physical or operational bounds restrict sensor displacement to any particular range, the distribution is inherent to the sampling process. Nevertheless, the skew towards smaller displacement values might influence the model's accuracy in underrepresented regions. 

\textit{6) Availability:} Dataset and code are publicly available at \href{https://github.com/airi-utcn/V-Beam-sensor}{V-Beam-sensor repository}.

\section{Exploratory trials}

\subsection{Random Forest Regressor}
The first trial consisted of using a multi-output Random Forest Regressor (scikit-learn) directly on the inverse problem, predicting $\beta$, l and w from the displacement, $\Delta T$, volume and stress. Initially the dataset consisted in 1000 samples. The motivation was that perhaps there was some linear relationship between the parameters that could be exploited. Even if multiple geometric parameters relate to the same input, maybe they would differ in stress and volume. Despite hyperparameter tuning via RandomizedSearchCV, performance was poor. The model always predicted constant values, failing to generalize over the whole input space.

\subsection{Expanded Dataset}
A second dataset was constructed with additional randomized sampling to improve the coverage of the parameter space (3000 samples). Both a Random Forest and an MLP were trained in an attempt to make the data more suitable for modeling. Systematic hyperparameter search was implemented, however, results remained inadequate, indicating a more profound structural issue, rather than insufficient data.

\subsection{Correlation Analysis}
\begin{figure}[htbp]
    \centering
    \includegraphics[width=0.8\columnwidth]{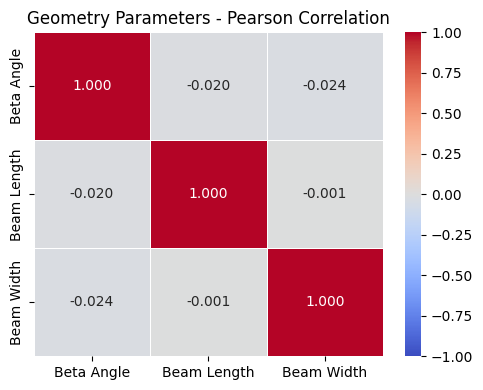}
    \caption{Pearson correlation matrix}
    \label{fig:matrix}
\end{figure}

To better understand the failure of the previous attempts, the third trial introduced systematic data analysis. Pearson, Kendall, and Spearman correlation matrices indicated a very weak correlation between  $\beta$, $l$ and $w$. Figure \ref{fig:matrix} shows the Pearson correlation matrix between the inputs of the model, as we can observe there is no relation we can deduce as the correlation is close to 0 for all pairs.

\begin{figure}[htbp]
    \centering
    \includegraphics[width=1\columnwidth]{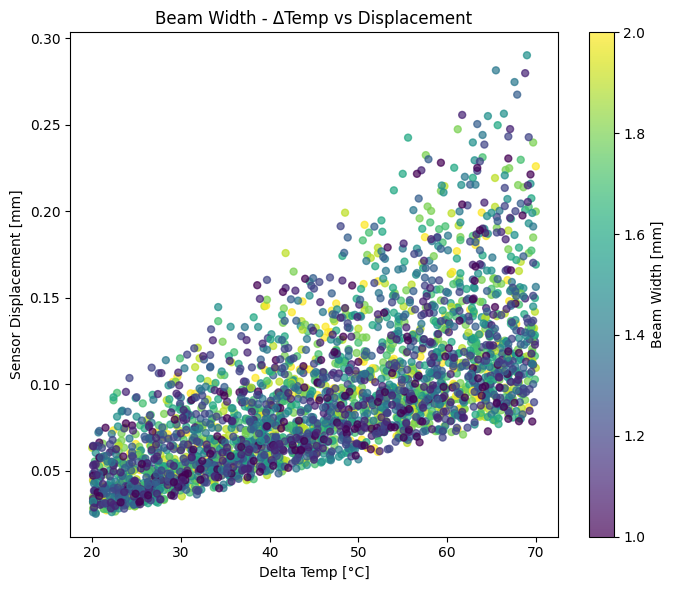}
    \caption{3D scatter plot of Beam width by $\Delta T$ and displacement}
    \label{fig:displacement_vs_DT}
\end{figure}

Three-dimensional scatter plots of $\Delta T$ and displacement colored by each geometric parameter revealed no structured mapping. For the same value of $\Delta T$ and displacement there were many values of $\beta$, l and w present. This confirmed the ill-posed nature of direct inverse regression and motivated a fundamental rethinking of the approach. Figure \ref{fig:displacement_vs_DT} shows the nature of the distribution of the beam with in relation with values of sensor displacement and $\Delta T$.

\subsection{Neural Network with a custom loss function}
\begin{figure}[htbp]
    \centering
    \includegraphics[width=1\columnwidth]{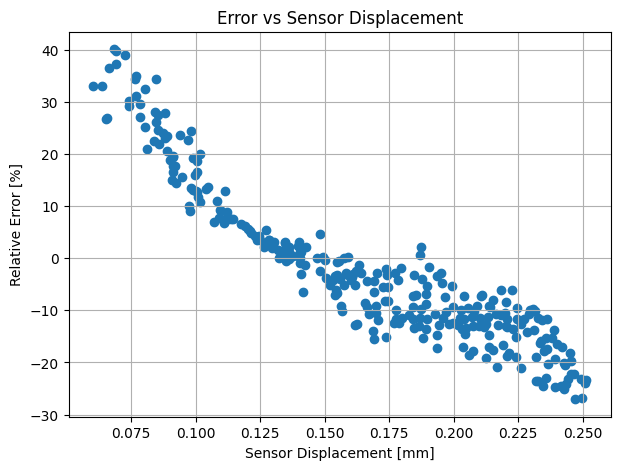}
    \caption{Error vs sensor displacement}
    \label{fig:errorNN}
\end{figure}

This set of trials focused on introducing a physics-informed loss function. We switched to PyTorch and a custom GeometryNet was trained to predict ($\beta$, l, w) with a composite loss combining:
\begin{itemize}
    \item MSE fit to true geometry
    \item a volume penalty computed analytically from predicted geometry
    \item a stress penalty computed from an analytical formula derived from the mechanical model
\end{itemize}
The idea was that, even if multiple geometries produce the same displacement, the physics penalties would guide the optimizer toward the minimum-stress, minimum-volume solution. We tried multiple combinations of weights for the components of the loss function. Increasing the weights for the stress and volume led to results where the predicted geometry had very small values for those two parameters but failed to get close to the target displacement. The best results came from a combination where the weights of the value and stress were only around 0.15 each, while the fit weight 0.7. This provided MAE of around 12\%. However, this result was deceptive as the NN always predicted a constant value (the mean) and because of the data distribution it seemed to give good results. Figure \ref{fig:errorNN} showcases exactly the error distribution mentioned earlier by plotting the error with regards to the displacement. 

\FloatBarrier 
\section{Two-Phase Workflow}
The failure of direct inverse regression motivated a two-phase strategy that sidesteps the ill-posedness by working in the forward direction.

\subsection{Analytical Formulas}

\begin{figure*}[htbp]
    \centering
    \includegraphics[width=\textwidth]{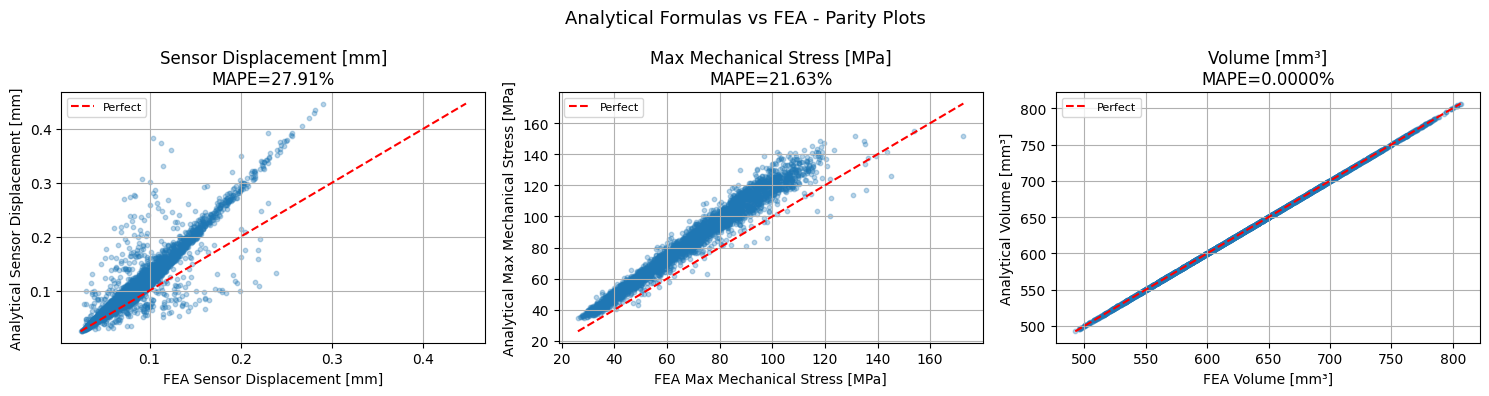}
    \caption{Analytical formulas analysis}
    \label{fig:analytical}
\end{figure*}

The analytical formulas presented in \cite{chiorean2014analytical} have a reported accuracy of under 6\%. However, testing has been performed on only 12 structures and this leads to the need of performing further research into the performance of the formulas. The results plotted in Figure \ref{fig:analytical} show a different truth. The MAPE of the formula for sensor displacement is of 27.91\%, for mechanical stress 21.63\%. The volume formula is, as expected, accurate. The results guide towards the usage of the formulas as a soft constraint in the loss function of the PINN, as the correct evolution trends of the parameters in relation to the sensor displacement are captured. 

\subsection{Forward Model}
A neural network forward model is trained to learn the well-posed physical mapping:
\begin{itemize}
    \item $f(\beta, l, w, \Delta T, E, \alpha) \mapsto (\text{Sensor Displacement, Max Stress, Volume})$
    \item E - Young's Modulus, $\alpha$ - Coefficient of Thermal Expansion 
\end{itemize}
Two types of forward models were trained on the same FEA dataset of 3{,}000 samples: MLP and PINN.

\begin{figure}[htbp]
  \centering
  \includegraphics[width=1\columnwidth]{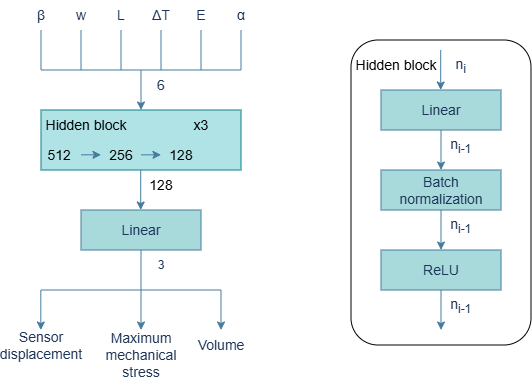}
  \caption{MLP\_FILTER Forward Neural Network Architecture}
  \label{fig:ann}
\end{figure}

Both types of models share the same fully-connected architecture defined by an identical class. However, each model preforms its own a grid search over a search space of 60 hyperparameter combinations. As a result, each model will have a different internal layout. For the MLP\_FILTER model, the winning configuration consists of hidden layers of widths 512, 256, 128. The first layer projects the 6 inputs into a 512 dimensional feature space and the final layer will map the feature space to 3 outputs. The architecture of this specific model can be seen in Figure \ref{fig:ann}. 

The hidden layer and hyperparameter configurations for each model are represented in Table \ref{tab:gss_configs}.

\begin{table}[h]
    \centering
    \caption{Optimal hyperparameter configurations selected by grid search.}
    \label{tab:gss_configs}
    \resizebox{\columnwidth}{!}{%
    \begin{tabular}{lcccc}
        \toprule
        Model        & Hidden layers         & Dropout & LR                 & Batch \\
        \midrule
        MLP          & $64 \to 64$           & 0.0     & $10^{-3}$          & 128 \\
        PINN         & $64 \to 64$           & 0.0     & $5 \times 10^{-4}$ & 128 \\
        MLP\_FILTER  & $512 \to 256 \to 128$ & 0.1     & $10^{-3}$          & 128 \\
        PINN\_FILTER & $128 \to 128$         & 0.0     & $10^{-3}$          & 128 \\
        \bottomrule
    \end{tabular}%
    }
\end{table}

All inputs and outputs were normalized with \texttt{StandardScaler} to equalize gradient magnitudes across variables
that span very different physical scales (e.g., $E = 71{,}000\ \text{MPa}$ alongside $\alpha = 23 \times 10^{-6}\
\text{K}^{-1}$). Training used the
Adam optimizer with $L_2$ weight decay ($10^{-4}$), gradient clipping (max norm 1.0), \texttt{ReduceLROnPlateau}
learning rate scheduling (factor 0.5, patience 15 epochs), and early stopping on validation loss (patience 30 epochs).

The chosen four forward models are:

\begin{itemize}
    \item \texttt{MLP} - this represents a simple multi layer perceptron trained on the whole unfiltered 3000 samples dataset. The loss function chosen for this model is MSE - trying to be as close as possible to the ground truth.
    \begin{equation}
        \text{MSE} = \frac{1}{n} \sum_{i=1}^{n} \left( y_i - \hat{y}_i \right)^2
    \end{equation}
    \item \texttt{MLP\_FILTER} - the same model as before, but this time training is performed on a filtered dataset and only includes samples that have $sensor displacement \in (0, 0.15)$. This is done in order to eliminate the outlier high-displacement values of the distribution.
    \item \texttt{PINN} - this model include a physics soft constrained to the overall loss of the network based on the physics formulas presented in \cite{chiorean2014analytical}. The loss function of the model is now:
    \begin{equation}
        \mathcal{L}_{\mathrm{PINN}}
        = \mathcal{L}_{\mathrm{data}}
        + \lambda\ \cdot \mathcal{L}_{\mathrm{physics}},
        \label{eq:pinn_loss}
    \end{equation}

    \noindent where
    
    \begin{align}
        \mathcal{L}_{\mathrm{data}}
            &= \frac{1}{N}\sum_{i=1}^{N}
               \bigl\lVert f_{\boldsymbol{\theta}}(\tilde{\mathbf{x}}_i)
               - \tilde{\mathbf{y}}_{\mathrm{FEA},i}\bigr\rVert^{2},
        \label{eq:l_data} \\[6pt]
        \mathcal{L}_{\mathrm{physics}}
            &= \frac{1}{N}\sum_{i=1}^{N}
               \bigl\lVert f_{\boldsymbol{\theta}}(\tilde{\mathbf{x}}_i)
               - \tilde{\mathbf{y}}_{\mathrm{analytical},i}\bigr\rVert^{2}.
        \label{eq:l_physics}
    \end{align} 

    \begin{itemize}
        \item $f_{\boldsymbol{\theta}}$ - this represents the neural network that has the learnable parameter $\theta$
        \item $\tilde{\mathbf{x}}_i$ - this represents the $i$-th input sample after the scaling process
        \item $\tilde{\mathbf{y}}_{\mathrm{FEA},i}$ - this is the standardized FEA simulation $i$-th sample output. It represents the ground truth for the prediction.
        \item $\tilde{\mathbf{y}}_{\mathrm{analytical},i}$ - this represents the target value computed according to the analytical formulas. It is a standardized values as well. 
        \item $\lambda$ - this is a parameter meant to signify the relative contribution of the physics term to the total loss. It must be a value greater than 0.\\
    \end{itemize}

    \item \texttt{PINN\_FILTER} - the same model as the one described above, but trained on the filtered dataset that \texttt{MLP\_FILTER} is also trained on
\end{itemize}

Each model must be evaluated using the same metric. The chosen metric is MAPE - mean absolute percentage error.

\begin{equation}
    \mathrm{MAPE} = \frac{100}{N} \sum_{i=1}^{N}
    \frac{\bigl| \hat{y}_i - y_i \bigr|}{\bigl| y_i \bigr|},
    \label{eq:mape}
\end{equation}

The dataset is split in training, validation and test sets following the 70\%/15\%/15\% rule. The performance of each forward model is analyzed using the test set. The results can be seen in Table \ref{tab:forward_model_mapeP}. The outcomes show that PINN results are better than MLP results. This is not a surprise as the physics constraints are expected to help guide the predictions towards physically feasible solutions. The only surprise is that the \texttt{MLP\_FILTER} model has a lower accuracy than the simple \texttt{MLP}. This is an indicator that filtering the data is not a universal solution for improving the performance of the models.

\begin{table}[ht]
\centering
\caption{Forward model test set performance across all four model variants.
         All values are Mean Absolute Percentage Errors (\%). Lower is better.}
\label{tab:forward_model_mapeP}
\begin{tabular}{lcccc}
\hline
\textbf{Model} &
\textbf{Disp. MAPE} &
\textbf{Stress MAPE} &
\textbf{Vol MAPE} \\
 & (\%) & (\%) & (\%) \\
\hline
MLP            & 8.48   & 2.19   & 0.42   \\
MLP\_FILTER    & 8.53   & 2.62   & 0.39   \\
PINN           & 7.84   & 1.95   & 0.35   \\
PINN\_FILTER   & 8.16   & 1.85   & 0.39   \\
\hline
\end{tabular}
\end{table}

\subsection{Gradient Descent Inverse Optimization}
After the training of the forward model, the weights of the NN are frozen and the inverse design is refrained as an optimization problem. The geometric parameters ($\beta$, l, w) are treated as learnable variables, initialized randomly from the training distribution, and updated by gradient descent to minimize a composite loss function. 
   \begin{equation}
    \mathcal{L} = \lambda_1 \cdot (\text{displ}_\text{pred} - \text{target})^2 + \lambda_2 \cdot \text{stress}_{pred} + \lambda_3 \cdot \text{volume}_{pred} 
   \end{equation} 
   \begin{equation}
       \lambda_1 = 1, \lambda_2 = 0.001, \lambda_3 = 0.001
   \end{equation}
To avoid stopping at a local minima, 10 independent restarts of the gradient descent are implemented, each performing up to 1000 iterations. The solution that outputs the smallest loss is selected. By this strategy, the error of the displacement prediction is near zero, while also identifying the solutions with minimum volume and stress. However, the MAPE score of almost 0\% is deceiving. The algorithm is able to always provide a configuration that according to the frozen model outputs the required displacement. However, the forward model is not perfect and thus propagates its error. The true test is to collect the geometries predicted by the gradient descent and use them as inputs for the FEA model in Ansys Mechanical. The results of the FEA reruns are visible in Table \ref{tab:results-mapeP}. The \texttt{PINN\_FILTER} is able to obtain the least amount of errors. 71\% of the predictions have under 5\% error, under 5\% being the golden standard in mechanical engineering. It can be seen that even though the \texttt{PINN} presented the best results on the forward model evaluation, the filtered version is better overall after applying the inverse design step. 

\begin{table}[ht]
\centering
\caption{Error percentages between the inverse design predictions and the FEA reruns}
\label{tab:results-mapeP}
\resizebox{\columnwidth}{!}{%
\begin{tabular}{lcccc}
\hline
\textbf{Model} &
\textbf{Disp. MAPE} &
\textbf{Median AE} &
\textbf{Within 1\%} &
\textbf{Within 5\%} \\
 & (\%) & (\%) & (\%) & (\%) \\
\hline
MLP          & 6.54 & 5.19 & 9.1  & 48.7 \\
MLP\_FILTER  & 7.56 & 4.36 & 12.4 & 54.6 \\
PINN         & 5.47 & 4.52 & 11.8 & 56.2 \\
PINN\_FILTER & 4.76 & 2.84 & 22.9 & 71.2 \\
\hline
\end{tabular}%
}
\end{table}

\FloatBarrier

The limitations of the current system come from the fact that a response is always given to the user without being able to express the true accuracy of the design. An inexperienced mechanical engineer could provide extreme combinations of $\Delta T$ and sensor displacement that are not physically possible. Nonetheless, the system will always provide an answer. This ML based workflow must be used by trained individuals that understand the possible associated risks.

\section{Conclusion}
The paper presents a two-phase machine learning framework for the inverse design of the V-Beam thermal sensor. This framework successfully solves the ill-posed nature of the inverse problem by training a physics-consistent forward neural network and embedding it within a gradient descent optimization loop. Five trials were needed in order to confirm the ill-posed nature of the direct inverse regression and adopt a different approach. The direct neural network with a custom loss function led to deceptive results. The final two-phase solution identifies optimal geometric parameters that match displacement targets while minimizing mechanical stress and material volume, offering a ML based approach to costly FEA simulations. There are several possible directions this work could take. The FEA dataset could be extended to include more values in the underrepresented zones of the distribution. This would target one of the main weaknesses of the process without having the need to resample the whole space. V-Beam sensor designs predicted by the workflow could be validated by the fabrication of physical sensors. This would confirm wether the reported accuracy holds true outside simulation. Augmenting the dataset with structures designed out of multiple materials and having a flexible number of beams would help with the overall generalization of the models. Future works could also focus on predicting the behavior of the V-Beam structure as an actuator instead of a sensor.

\bibliographystyle{IEEEtran}
\bibliography{bib} 

\nocite{cai2021physics}
\nocite{raissi2019physics}
\nocite{lin2026machine}
\nocite{lee2023machine}
\nocite{popescu2009multilayer}
\nocite{adler2017solving}
\nocite{engl2014inverse}
\nocite{ruder2016overview}
\nocite{ballard2021machine}
\nocite{rodriguez2015machine}
\nocite{ebert2021cira}
\nocite{jagota2013finite}
\nocite{guo2021artificial}
\nocite{kabanikhin2008definitions}
\nocite{argoul2012overview}
\nocite{golub11998tikhonov}
\nocite{peurifoy2018nanophotonic}
\nocite{alzubaidi2023survey}
\nocite{hoffer2021mesh}
\nocite{sun2019survey}
\nocite{dozat2016incorporating}
\nocite{chan2025inverse}
\nocite{huang2015proposed}
\nocite{beeby2004mems}
\nocite{choudhary2017mems}
\nocite{mukherjee2003mems}
\nocite{chiorean2014deflection}
\nocite{thompson2017ansys}
\nocite{cozad2014learning}
\nocite{emworks_vbeam}
\nocite{kaufman2012leakage}
\nocite{kingma2014adam}
\nocite{lecu7022006alloygraph}

\end{document}